\documentclass[aps,prl,twocolumn,amssymb,amsmath,floatfix]{revtex4}
\usepackage{graphicx,subfigure}
\usepackage{bm}
\usepackage{verbatim}
\usepackage{amsmath}
\usepackage{amssymb}
\usepackage[T1]{fontenc}
\usepackage{ae,aecompl}

\newcommand{\curH}{{\cal H}}

\newcommand{\bk}{{\bf k}}

\newcommand{\grad}{{\bm{\nabla}}}

\newcommand{\br}{{\bf r}}

\newcommand{\be}{\begin{equation}}
\newcommand{\ee}{\end{equation}}
\newcommand{\bea}{\begin{eqnarray}}
\newcommand{\eea}{\end{eqnarray}}
\newcommand{\ve}{\vert}

\def\rf#1{(\ref{#1})}
\def\rfs#1{Eq.~\rf{#1}}

\begin{document}
\title{$p$-Wave Resonant Bose Gas: A Finite-Momentum Spinor Superfluid}
\author{Leo Radzihovsky}
\author{Sungsoo Choi}
\affiliation{
Department of Physics, 
University of Colorado, 
Boulder, CO 80309}
\date{\today}
\begin{abstract}
  We show that a degenerate gas of two-species bosonic atoms
  interacting through a $p$-wave Feshbach resonance (as realized
  in, e.g., a $^{85}$Rb-$^{87}$Rb mixture) exhibits a {\it finite-momentum} 
  atomic-molecular superfluid (AMSF), sandwiched by a molecular $p$-wave 
  (orbital spinor) superfluid and by an $s$-wave atomic superfluid at large
  negative and positive detunings, respectively. The magnetic
  field can be used to tune the modulation wave vector of the AMSF
  state, as well as to drive quantum phase transitions in this rich
  system.
\end{abstract}

\maketitle
A Feshbach resonance (FR) is an exceptionally fruitful experimental
``knob'' that allows exquisite tunability of interactions in
degenerate atomic gases. This has led to realizations and studies of
Bose-Einstein condensation (BEC)-BCS crossover of fermion-paired $s$-wave
superfluidity~\cite{regal04,zwierlein04,kinast04,chin_grimm04}. The
bosonic counterparts have also been extensively explored, and in fact,
in the $s$-wave FR case, e.g., in $^{85}$Rb~\cite{wieman00}, predate
recent fermionic developments.  As was recently
emphasized~\cite{radzihovsky_boson04,radzihovsky_boson08,romans04},
in contrast to their fermionic analogs that undergo a smooth BEC-BCS
crossover, resonant bosonic gases are predicted to exhibit magnetic
field and/or temperature driven sharp phase transitions between
distinct molecular and atomic superfluid phases.

Motivated by these successes, recent attention has focused on a
realization of an even richer $p$-wave paired fermionic
superfluidity~\cite{HoPwave,GRAprl,GRaop,yip05},
utilizing $p$-wave FR in $^{40}$K and $^6$Li.  Laboratory production
of $p$-wave Feshbach molecules~\cite{gaebler07,zhang_salomon04}
showed considerable promise toward this goal; however, reaching
molecular degeneracy has been plagued with short molecular
lifetimes.

In another important development, experiments on a $^{85}$Rb-$^{87}$Rb
mixture have demonstrated a $p$-wave FR at $B=257.8$G
between these two bosonic isotopes~\cite{papp08}.
Although consequences of this two-body $p$-wave resonance on the
degenerate state of such a gas mixture has not been further explored
experimentally, it provides the main motivation for our work. In this
Letter, we report on our study of a two-species degenerate Bose gas with a $p$-wave Feshbach
resonant interspecies interaction. 

\begin{figure}[thb]
\includegraphics[width=8.25cm]{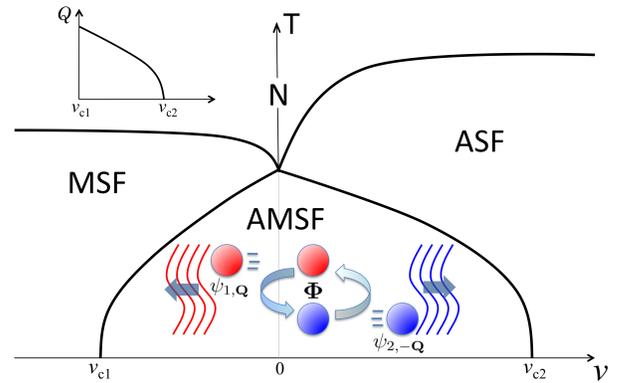}
\caption{(color online). Schematic temperature-detuning phase diagram for a
  two-species mixture of bosonic atoms exhibiting atomic (ASF),
  molecular (MSF), and atomic-molecular (AMSF) phases. In the AMSF
  $p$-wave, molecular condensate coexists with a finite-momentum $Q$
  (see left inset) atomic superfluid. The cartoon in
  AMSF phase illustrates the mechanism driving finite-momentum
  condensation.}
\label{phasediagram}
\end{figure}

As summarized by the phase diagram in
Fig.~\ref{phasediagram}, we find (within a mean-field treatment that
we expect to largely survive fluctuations) that in addition to the
normal (N, i.e., nonsuperfluid) phase, the $p$-wave Feshbach resonant
two-component {\em balanced} Bose gas (e.g., equal mixture of
$^{85}$Rb and $^{87}$Rb atoms) exhibits three classes of superfluid
phases: atomic (ASF), molecular (MSF), and atomic-molecular (AMSF)
condensates. Our most interesting finding is that the AMSF, sandwiched
between (large positive detuning) ASF and (large negative detuning)
MSF phases is necessarily 
a {\it finite-momentum Q} spinor superfluid,
with a characteristic wave vector (with $\hbar=1$)
\begin{equation}
Q=\alpha m \sqrt{n_m} \sim \sqrt{\gamma_p\ell n_m}\lesssim\sqrt{\gamma_p}/\ell,
\label{Q}
\end{equation}
tunable with a magnetic field (via FR detuning, $\nu$ that primarily
enters through the molecular condensate density $n_m(\nu)$), with
$\alpha$, $m$, $\ell$, and $\gamma_p$, respectively, the FR
coupling, atomic mass, atom spacing, and
a dimensionless measure of FR width\cite{GRaop}.

Within the narrow ($\gamma_p \ll 1$) FR
approximation\cite{GRAprl,radzihovsky_boson04,GRaop,radzihovsky_boson08},
we find that the optimum collinear state is characterized by a single
${\bf Q}$, Fulde-Ferrell-like state~\cite{ff64}, as opposed to a
$+{\bf Q}$ and $-{\bf Q}$ Larkin-Ovchinnikov-like state\cite{lo65}
found in imbalanced paired fermionic
systems\cite{MiMaI,SheehyRadzihovskyPRL,SheehyRadzihovskyAOP}.
However, a more detailed study is necessary to ascertain the precise
set of $Q$'s at which AMSF condensation takes place.

The physical picture behind the finite-momentum AMSF formation is
illustrated in Fig.~\ref{phasediagram}. At intermediate
detuning where atomic gap closes within the MSF state, $p$-wave
molecules decay via FR into a pair of atoms, which (due to the
$p$-wave nature of the molecules) are necessarily created at finite
and opposite momenta, $\pm\bk$, and therefore at low temperature form
a finite-momentum atomic condensate, AMSF. The energetic cost ($\sim
k^2/2m$) of a finite momentum atomic condensation is balanced by the
lowering of the energy ($\sim\alpha k\sqrt{n_m}$) through FR
hybridization between closed-channel $p$-wave molecule and
open-channel pair of atoms that is only possible at finite atomic
momentum $k$, giving $Q$ in \rfs{Q}.

As we detail below, in addition, we find that each of the three
superfluid classes (ASF, MSF, AMSF) in turn consist of distinct phases
selected by detuning, temperature, and background $s$-wave scattering
lengths, and distinguished by the nature (ferromagnetic or polar) of
the $p$-wave molecular condensate and/or which combination of the two
types of atoms is Bose condensed. We explore the nature of these SF
phases and associated phase transitions.

To outline the derivation of these results, we consider a model of a
gas mixture of two distinguishable bosonic atoms (e.g., $^{85}$Rb,
$^{87}$Rb)~\cite{papp08}, created by field operators
$\hat\psi_\sigma^\dagger(\br)
=\left(\hat\psi_1^\dagger(\br),\hat\psi_2^\dagger(\br)\right)$,
and interacting through a $p$-wave Feshbach resonance associated with a
tunable closed-channel bound state. The corresponding $p$-wave
($\ell=1$) closed channel hetero-molecule (e.g., $^{85}$Rb-$^{87}$Rb) is created
by a vector field operator
$\hat{\bm\phi}^\dagger(\br)
=(\hat\phi_x^\dagger,\hat\phi_y^\dagger,\hat\phi^\dagger_z)$.
This system is governed by a grand-canonical Hamiltonian
$H[\hat{\psi}_\sigma,\hat{\bm\phi}]=\int d^3r\curH$, with
\bea
\curH &=&  
\sum_{\sigma=1,2}\hat\psi_{\sigma}^{\dag} \hat{\varepsilon}_{\sigma}
\hat\psi_\sigma 
+\hat{\bm\phi}^{\dag}\cdot\hat{\omega}\cdot\hat{\bm\phi} 
+ \curH_{bg} \label{H} \\
&&+\frac{\alpha}{2}\hat{\bm\phi}^{\dag}\cdot
\left[\hat\psi_{1}(-i\grad)\hat\psi_{2} -
\hat\psi_{2}(-i\grad)\hat\psi_{1}\right]+H.c.,\nonumber
\eea
where
$\hat{\varepsilon}_{\sigma} =-\frac{1}{2m}\grad^2-\mu_\sigma$,
$\hat{\omega}=-\frac{1}{4m}\grad^2-\mu_m$ with the molecular
chemical potential $\mu_m=\mu_1+\mu_2-\nu$ adjustable by
detuning $\nu$. We have taken atomic
masses to be identical (a good approximation for the $^{85}$Rb-$^{87}$Rb mixture)~\cite{commentFRequalmass}, 
and focus on the balanced case of
$\mu_1=\mu_2=\mu$, with $\mu$ fixing the total
number of $^{85}$Rb and $^{87}$Rb atoms, whether in the (open-channel) atomic or
(closed-channel) molecular form.

The FR interaction encodes a coherent interconversion between a pair
of open-channel atoms $1,2$ (in a singlet combination of $1,2$ labels,
as required by bosonic statistics) and a closed-channel $p$-wave
molecule, with amplitude $\alpha$\cite{commentFRequalmass}.
We have focused on a rotationally invariant FR
interaction, with $\hat{\omega}$ and $\alpha$ independent of the
molecular component $i$. This is an approximation to $^{85}$Rb-$^{87}$Rb mixture,
where the $p$-wave FR around $B=257.8$G is split into a doublet
by $\Delta B=0.6$G, similarly to the fermionic case of
$^{40}$K\cite{RegalPwave,gaebler07,GRAprl,GRaop}. We leave the more realistic,
richer case for future studies\cite{ChoiRadzihovskyPRA}.

The FR coupling $\alpha$ and detuning $\nu$ are fixed experimentally
through measurements of the low-energy $p$-wave scattering
amplitude\cite{RegalPwave,gaebler07} $f_p(k) =
\frac{k^2}{-v^{-1}+\frac{1}{2}k_0k^2-ik^3}$. The scattering volume,
$v$ (tunable via $\nu$) and the
characteristic wave vector, $k_0$ (a $p$-wave analog of 
the effective range, negative for the FR case)~\cite{GRaop} 
define our model parameters in terms of these
experimental observables.

The background (nonresonant) interaction density
$\curH_{bg}=\curH_{a} + \curH_{m} + \curH_{am}$ consists of
\bea
\curH_a &=&\sum_{\sigma=1,2}\frac{\lambda_\sigma}{2}\hat\psi^{\dag
    2}_{\sigma}\hat\psi^{2}_{\sigma}
+ \lambda_{12}\hat\psi^{\dag}_{1}\hat\psi^{\dag}_{2}\hat\psi_{2}\hat\psi_{1}, \\
\curH_m
&=&\frac{g_1}{2}(\hat{\bm\phi}^{\dag}\cdot\hat{\bm\phi})^2
+\frac{g_2}{2}\ve\hat{\bm\phi}\cdot\hat{\bm\phi}\ve^2, \\
\curH_{am} &=&\sum_{\sigma=1,2} g_{am}
\hat\psi^{\dag}_{\sigma}\hat{\bm\phi}^\dag\cdot\hat{\bm\phi}\hat\psi_{\sigma},
\eea
where coupling constants $\lambda_\sigma$, $\lambda_{12}$, $g_{1,2}$,
$g_{am}$ are related to the corresponding $s$-wave scattering lengths
($a_1$, $a_2$, etc.)  in a standard way, and thus are fixed
experimentally through measurements on the gas in a dilute limit. The
miscibility of a two-component atomic gas requires\cite{Greene} $a_1
a_2 > a_{12}^2$, which may be problematic for the case of
$^{85}$Rb-$^{87}$Rb due to the negative background scattering length
of $^{85}$Rb.

The molecular interaction couplings $g_1$, $g_2$ (set by the $L=0$ and
$L=2$ channels of $p$-wave molecule-molecule scattering) and $g_{am}$
can be derived from a combination of $s$-wave atom-atom
($\lambda_\sigma$) and $p$-wave FR ($\alpha$)
interactions\cite{ChoiRadzihovskyPRA}.

Qualitative features of the phase diagram for the system can be mapped
out through a mean-field treatment of the Hamiltonian \rf{H}. This
amounts to a minimization of the Landau free-energy functional
$F[\Psi_\sigma, {\bm\Phi}]$ of classical fields $\Psi_\sigma(\br)$,
${\bm\Phi}(\br)$, corresponding to the coherent state field
configurations for the atomic and molecular operators. In the simplest
approximation, $F[\Psi_\sigma, {\bm\Phi}]$ takes the form identical to
$H[\hat{\psi}_\sigma,\hat{\bm\phi}]$, with the effective couplings
($\tilde{\mu}_\sigma, \tilde{\mu}_m, \tilde{\lambda}_\sigma,\ldots$)
that are functions of microscopic parameters ($\mu_\sigma, \nu,
\lambda_\sigma,\ldots$) appearing in \rf{H}.

Minimization of $F[\Psi_\sigma, {\bm\Phi}]$ is quite
straightforward\cite{ChoiRadzihovskyPRA}. For large {\em positive}
detuning $\nu$, closed-channel molecules are gapped, and the ground
state is a molecular vacuum. Thus, at low temperature $\tilde{\mu}_m <
0$ and $F[\Psi_\sigma, {\bm\Phi}]$ is minimized by ${\bm\Phi}=0$,
reducing to $F_a[\Psi_\sigma]=F[\Psi_\sigma, 0]=\int d^3
r\left[\sum_{\sigma=1,2}\left(\Psi_{\sigma}^*\hat{\tilde{\varepsilon}}_{\sigma}
    \Psi_\sigma +
    \frac{\tilde{\lambda}_\sigma}{2}|\Psi_\sigma|^4\right) +
  \tilde{\lambda}_{12}|\Psi_{1}|^2|\Psi_{2}|^2\right]$. This
functional is a special ($U(1)\times U(1)$) case of a $O(N)\times
O(M)$ model that has been studied
extensively\cite{LiuFisher,KNF,Vicari}. The free-energy is clearly
minimized by uniform $\Psi_1,\Psi_2$, as this lowers atomic kinetic
energy. For $\tilde{\lambda}_1\tilde{\lambda}_2 > \tilde{\lambda}_{12}^2$ in addition to
the normal (nonsuperfluid) state, the system exhibits three ASF
phases: (i) ASF$_1$ with $\Psi_1\neq 0,\Psi_2 = 0$, (ii) ASF$_2$ with
$\Psi_1= 0,\Psi_2 \neq 0$, (iii) ASF$_{12}$ with $\Psi_1\neq 0,\Psi_2
\neq 0$, separated by continuous phase transitions.  For a balanced
mixture $\tilde{\mu}_1 = \tilde{\mu}_2$, the system exhibits a direct
N-ASF$_{12}$ transition through a tetracritical point, $\tilde{\mu}_1
= \tilde{\mu}_2 = 0$, that is believed to be in the decoupled
universality class\cite{LiuFisher,KNF,Vicari}. For
$\tilde{\lambda}_1\tilde{\lambda}_2 < \tilde{\lambda}_{12}^2$, the ASF$_{12}$ phase is
absent, and ASF$_{1}$ and ASF$_{2}$ are separated by a first-order
transition that terminates at a bicritical
point\cite{LiuFisher,KNF,Vicari}. All other transitions
(N-ASF$_1$, N-ASF$_2$, and ASF$_i$-ASF$_{12}$) are in the $XY$
universality class, breaking associated $U(1)$ symmetries. The phase
boundaries and the values of the atomic condensate order parameters
can be straightforwardly computed within mean-field
theory (MFT)\cite{ChoiRadzihovskyPRA}, but are modified by
fluctuations\cite{KNF,Vicari}.

Within ASF phases, the spectrum of fluctuations can be
straightforwardly computed by a Bogoliubov diagonalization of coupled
atomic and molecular excitations, with details depending on which of
the three possible ASF phases is studied. In general, there will be one
Bogoliubov sound mode per broken atomic $U(1)$ symmetry, with one
Goldstone mode in ASF$_1$ and ASF$_2$ phases and two in
ASF$_{12}$\cite{ChoiRadzihovskyPRA}.

In the opposite limit of large {\em negative} detuning, $\nu$
open-channel atoms are gapped, and the ground state is an atomic
vacuum. Hence, at low temperature $\tilde{\mu} < 0$ and $F[\Psi_\sigma,
{\bm\Phi}]$ is minimized by $\Psi_\sigma=0$ and a uniform molecular
condensate ${\bm\Phi}$, reducing to $F_m[{\bm\Phi}]/V=F[0,
{\bm\Phi}]/V= -\tilde{\mu}_m|{\bm\Phi}|^2
+\frac{\tilde{g}_1}{2}({\bm\Phi}^*\cdot{\bm\Phi})^2
+\frac{\tilde{g}_2}{2}\ve{\bm\Phi}\cdot{\bm\Phi}\ve^2$, for (orbital $\ell=1$)
spin-$1$ molecular bosons with well-studied thermodynamics~\cite{HoSpinor,Ohmi,StamperKurn,Cornell}.  
In particular, we
predict our system to also exhibit polar (MSF$_{\text{p}}$ for
$\tilde{g}_2 < 0$) and ferromagnetic (MSF$_{\text{fm}}$ for
$\tilde{g}_2 > 0$) molecular condensates, respectively, corresponding
to ${\bm\Phi} = \Phi_0\hat{\bm n}\in [S_2 \times U_N(1)]/Z_2$ and
${\bm\Phi} = \Phi_0(\hat{\bm n}+i\hat{\bm m})/\sqrt{2}\in SO(3)$ order
parameters, with $\hat{\bm n},\hat{\bm m}, \hat{\bm
  \ell}\equiv\hat{\bm n}\times\hat{\bm m}$ an orthonormal triad and
$\Phi_0$ a complex amplitude, breaking $SO(3)\times
U_N(1)$\cite{Mukerjee,commentU_N}. The finite $T$ N-MSF transitions
are in the universality class of a complex $O(3)$ model\cite{Vicari}.

The Goldstone mode content of these orbital $\ell=1$ molecular
condensate phases is also identical to that of spinor
condensates\cite{HoSpinor}, with MSF$_{\text{p}}$ exhibiting three
$E_m^{(\text{MSF}_{\text{p}})}(k)\sim k$ Bogoliubov modes and MSF$_{\text{fm}}$ characterized
by one $E_{m1}^{(\text{MSF}_{\text{fm}})}(k)\sim k$ Bogoliubov and one
$E_{m2}^{(\text{MSF}_{\text{fm}})}(k)\sim k^2$ ferromagnetic spin-wave modes.  An
attractive new feature of these orbital molecular condensates
is that $\tilde{g}_2(\nu)$ is a
tunable function of detuning that can be used to drive
a 1st order MSF$_{\text{p}}$-MSF$_{\text{fm}}$ transition\cite{GRaop,ChoiRadzihovskyPRA}.

To calculate the spectrum of low-energy excitations inside the MSF
(polar and ferromagnetic) phases, we separate molecular field
$\hat{\bm\phi}={\bm\Phi}+\hat{\bm\varphi}$ into a condensate and small
fluctuations about it, obtaining
$\curH[\hat{\psi}_\sigma,{\bm\Phi}+\hat{{\bm\varphi}}]\approx {\cal
  E}_{g}^{(0)}[{\bm\Phi}] + \curH_a^{(2)}+\curH_m^{(2)}$, where ${\cal
  E}_{g}^{(0)}[{\bm\Phi}]=\curH[0,{\bm\Phi}]$ is the zeroth-order
approximation to the MSF ground state energy,
\begin{eqnarray}
\curH_a^{(2)}&=&
\sum_{\sigma=1,2}\hat\psi_{\sigma}^{\dag} \tilde{\varepsilon}_{\sigma}
\hat\psi_\sigma
+\alpha{\bm\Phi}\cdot\hat\psi_{1}(-i\grad)\hat\psi_{2} +H.c.
\label{Ha2}
\end{eqnarray}
is the quadratic atomic Hamiltonian density, and $\curH_m^{(2)}
=\hat{\varphi}_i^{\dag}\tilde{\omega}_{ij}\hat{\varphi}_j
+\frac{g_1}{2}\Phi_i^*\Phi_j^*\hat{\varphi}_i\hat{\varphi}_j
+\frac{g_2}{2}{\bm\Phi}^*\cdot{\bm\Phi}^*\hat{\bm\varphi}\cdot\hat{\bm\varphi}
+ H.c.$ is the quadratic molecular Hamiltonian density, with
$\tilde{\varepsilon}_{\sigma}=\hat{\varepsilon}_{\sigma}+g_{am}|{\bm\Phi}|^2$
and $\tilde{\omega}_{ij}=(\hat{\omega}+g_1|{\bm\Phi}|^2)\delta_{ij} +
g_1{\Phi}^*_j{\Phi}_i+ 2g_2{\Phi}^*_i{\Phi}_j$.

To this quadratic order, the molecular and atomic excitations decouple
and can therefore be diagonalized independently. The molecular part
has been extensively studied in the context of spinor $F=1$
condensates\cite{HoSpinor}. For the polar ($g_2 < 0$) MSF$_{\text{p}}$ state,
there are three ``sound'' modes, one Bogoliubov type with sound
velocity $c_{||}^{(\text{MSF}_{\text{p}})}=\sqrt{(g_1+g_2)n_m/2m}$, and other two
degenerate spin-waves with velocity
$c_{\perp}^{(\text{MSF}_{\text{p}})}=\sqrt{|g_2|n_m/2m}$.  For the ferromagnetic
($g_2 > 0$) MSF$_{\text{fm}}$ state, there is one Bogoliubov mode, with sound
velocity $c^{(\text{MSF}_{\text{fm}})}=\sqrt{g_1n_m/2m}$, one quadratic $k^2/2m$
ferromagnetic spin-wave mode, and one gapped quadratic $k^2/2m +
2g_2n_m$ canonically conjugate mode.

The atomic sector can also be readily diagonalized, giving
\begin{eqnarray}
E_{(a)}^{(\text{MSF})}(k)
&=&\sqrt{\tilde{\varepsilon}_k^2-\alpha^2|{\bm\Phi}\cdot\bk|^2},
\end{eqnarray}
where $\tilde{\varepsilon}_k = k^2/2m - \mu + g_{am} n_m$ and $n_m$ is
the molecular condensate density. The details of the spectrum only
differ quantitatively between the MSF$_{\text{p}}$ and
MSF$_{\text{fm}}$ phases, both exhibiting a minimum at a {\em finite}
${\bf k}_{\text{min}}$ for $\nu > \nu_*$, and an atomic gap
$\Delta(\nu)\equiv E_{(a)}^{(\text{MSF})}(k_{\text{min}})$ that closes at the
transition $\nu_{c_1}$ into the corresponding AMSF states.  Such
spectrum (displaying a finite $k$ minimum) should be experimentally
observable through, e.g., rf spectroscopy and is also a complementary
way to detect the approaching finite $Q$ instability (phase
transition) into the AMSF state.  Simple analysis inside
MSF$_{\text{p}}$ gives
\begin{eqnarray}
{\bf k}_{\text{min}}^{(\text{MSF}_{\text{p}})}&=&\hat{\bf n}\sqrt{(2m^2\alpha^2+m g)n_m +
  m\nu},\nonumber\\
\Delta^{(\text{MSF}_{\text{p}})} &=&\sqrt{-(m\alpha^2+g)m\alpha^2n_m^2 
 - m\alpha^2\nu n_m},\nonumber\\
\nu_{*}^{(\text{MSF}_{\text{p}})} &=& -(2m\alpha^2+g)n_m,\nonumber\\
\nu_{c_1}^{(\text{MSF}_{\text{p}}-\text{AMSF}_{\text{p}})}&=& -(m\alpha^2+g)n_m,
\end{eqnarray}
with $g=g_1+g_2-2g_{am}$ and we used lowest order MSF$_{\text{p}}$ relation
$\mu_m\equiv 2\mu-\nu \approx (g_1+g_2)n_m$ to eliminate the atomic
$\mu$ in favor of molecular condensate $n_m$ and
detuning $\nu$. The corresponding expressions inside MSF$_{\text{fm}}$ differ
only slightly\cite{ChoiRadzihovskyPRA}, except that ${\bf
  k}_{\text{min}}^{(\text{MSF}_{\text{fm}})}$ lies in the $\hat{\bm n}$-$\hat{\bm m}$ plane
perpendicular to the ferromagnetic quantization axis, $\hat{\bm \ell}$
rather than along it as in the MSF$_{\text{p}}$ state.

Upon further increase of $\nu$, the atomic gap
$\Delta^{(\text{MSF})}(\nu)$ closes at $\nu_{c_1}$ and atoms
Bose-condense at a finite ${\bf k}_{\text{min}}^{(\text{MSF})}(\nu_{c_1})$
(${\bf k}_{\text{min}}^{(\text{MSF}_{\text{p}})} =m\alpha\sqrt{n_m}$, ${\bf
  k}_{\text{min}}^{(\text{MSF}_{\text{fm}})}=m\alpha\sqrt{n_m}/\sqrt{2} $),
thereby breaking the remaining $Z_2\times U_{\Delta N}(1)$ symmetry,
with $Z_2$ corresponding to the discrete part of (atom number)
$U_N(1)$ unbroken in the paired MSF states. The two associated order
parameters are given by $\Psi_\pm(\br)=\sum_{{\bf
    Q}_n}\left(\pm\Psi_{{\bf Q}_n,1}e^{i({\bf Q}_n\cdot\br-\theta_{\bf
      Q})} +\Psi^*_{-{\bf Q}_n,2}e^{i{\bf Q}_n\cdot\br}\right)$, where
$\theta_{\bf Q}$ is the phase of $\Delta_{\bf
  Q}\equiv\alpha{\bm\Phi}\cdot{\bf Q} =|\alpha{\bm\Phi}\cdot{\bf
  Q}|e^{i\theta_{\bf Q}}$ and ${\bf Q}_n = n {\bf Q}$ ($n\in {\cal
  Z}$, an integer). The critical transition point for $\Psi_+$ and
$\Psi_-$ is split by $\pm|\Delta_{\bf Q}|$, respectively, so only
$\Psi_-$ condenses at $\nu_{c_1}$, allowing for the possibility
of two transitions, MSF-AMSF$_-$ followed by AMSF$_-$-AMSF$_+$. 
The values of order parameters and transition
points can be straightforwardly worked out. Within MFT, we find that a
single-$Q$ state is energetically preferred\cite{ChoiRadzihovskyPRA},
but we do not expect this to survive generically.

In addition to the molecular and atomic superfluidity,
AMSF$^{\text{p},\text{fm}}_{+,-}$ phases are finite-momentum condensates that
generically can exhibit crystalline
order and therefore are supersolids (at least in
the case of more than one ${\bf Q}_n$
condensation\cite{singleQcomment}). For a collinear set of ${\bf
  Q}_n$'s, the supersolid is a unidirectional density wave breaking
translational invariance along ${\bf Q}_n$, with the latter aligned with
the quantization axis, $\hat{\bm n}$ of the MSF$_{\text{p}}$ state and
transverse to the quantization axis $\hat{\bm\ell}$ (i.e., lying in
the $\hat{\bm n}$-$\hat{\bm m}$ plane) of the MSF$_{\text{fm}}$
state. Thus, while rotational $O(2)$ symmetry about the $\hat{\bm n}$
axis remains intact inside the AMSF$_{\text{p}}$ state, it is
spontaneously broken by such a uniaxial density wave inside the
AMSF$_{\text{fm}}$ state. It is notable that in this latter case, in
the presence of fluctuations, such superfluid density wave will
exhibit quantum liquid-crystal phenomenology similar to that of the
fermion-paired Larkin-Ovchinnikov
superfluid\cite{RadzihovskyVishwanathPRL,ChoiRadzihovskyPRA}. The
excitation spectra inside AMSF phases can be computed via a
generalized Bogoliubov transformation and involve a
diagonalization of a $10\times 10$ matrix (corresponding to three and
one coupled complex molecular and atomic fields), 
leading to Goldstone modes consistent with above
symmetry-based arguments\cite{ChoiRadzihovskyPRA}.

The nature of MSF$_{\text{p},\text{fm}}$-AMSF$_{\text{p},\text{fm}}$
transitions (beyond MFT) remains an open question. Based on the
experience with spinor condensates\cite{Mukerjee} and LO
superfluid\cite{RadzihovskyVishwanathPRL}, we expect this system to
exhibit a variety of fractional composite topological defects. We
leave detailed study of these to a future
publication\cite{ChoiRadzihovskyPRA}.

Upon further increase of $\nu$, the AMSF transitions into the
ASF at $\nu_{c2}\approx(2\lambda-g_{am})n_a$ 
(with $\lambda = \frac{1}{4}(\lambda_1+\lambda_2+2\lambda_{12})$ and the atomic condensate density $n_a$), 
determined by the point of
vanishing of the $n_m(\nu)$ and $Q(\nu)$. The AMSF can thus exhibit a broad
stability range $\nu_{c2}-\nu_{c1}$,
set by a combination of $n_a, n_m, \lambda_i,g_i,...$,  
that in 3d we expect to
survive beyond our mean-field analysis.

To summarize, we studied a degenerate gas of two-species bosonic atoms
interacting through a $p$-wave Feshbach resonance, as realized in a
$^{85}$Rb-$^{87}$Rb mixture. We showed that at intermediate FR detuning, such gas
exhibits an atomic-molecular superfluid (AMSF) state condensed at a
finite momentum, that undergoes phase transitions into a molecular
$p$-wave (orbital spinor) superfluid (MSF) and into an $s$-wave atomic
superfluid (ASF) at large negative and positive detunings,
respectively. A magnetic field can be used to tune the modulation
wave vector of the AMSF between zero and a value set by interactions as
well as to drive quantum phase transitions in this rich system.

We thank V. Gurarie for discussions and acknowledge
support by the NSF No. DMR-0321848 (L. R., S. C.), Berkeley Miller, and
University of Colorado Faculty Fellowships (L. R.).  L. R. thanks Berkeley
Physics Department for its hospitality during part of this work.

\begin{thebibliography}{99}
\bibitem{regal04} C. A. Regal
{\it et al.}, Phys. Rev. Lett. {\bf 92}, 040403 (2004).
\bibitem{zwierlein04} M. W. Zwierlein
{\it et al.}, Phys. Rev. Lett. {\bf 92}, 120403 (2004).
\bibitem{kinast04} J. Kinast
{\it et al.}, Phys. Rev. Lett. {\bf 92}, 150402 (2004).
\bibitem{chin_grimm04} C. Chin
{\it et al.}, Science {\bf 305}, 1128 (2004).
\bibitem{wieman00} S. L. Cornish
{\it et al.}, Phys. Rev. Lett. {\bf 85}, 1795 (2000).
\bibitem{radzihovsky_boson04} L. Radzihovsky
{\it et al.}, Phys. Rev. Lett. {\bf 92}, 160402 (2004).
\bibitem{radzihovsky_boson08} L. Radzihovsky
{\it et al.}, Ann. Phys. (N. Y.) {\bf 323}, 2376 (2008).
\bibitem{romans04} M. W. J. Romans
{\it et al.}, Phys. Rev. Lett. {\bf 93}, 020405 (2004).
\bibitem{HoPwave} T.-L. Ho and R. B. Diener, 
Phys. Rev. Lett. {\bf 94}, 090402 (2005).
\bibitem{GRAprl} V. Gurarie
{\it et al.}, Phys. Rev. Lett. {\bf 94}, 230403 (2005).
\bibitem{GRaop} V. Gurarie and L. Radzihovsky, 
Ann. Phys. (N. Y.) {\bf 322}, 2 (2007).
\bibitem{yip05} C. Cheng and S.-K. Yip, Phys. Rev. Lett. {\bf 95}, 
070404 (2005).
\bibitem{gaebler07} J. P. Gaebler
{\it et al.}, Phys. Rev. Lett. {\bf 98}, 200403 (2007).
\bibitem{zhang_salomon04} J. Zhang
{\it et al.}, Phys. Rev. A {\bf 70}, 030702 (2004).
\bibitem{papp08} S. B. Papp
{\it et al.}, Phys. Rev. Lett. {\bf 101}, 040402 (2008).
\bibitem{ff64} P. Fulde and R. A. Ferrell, Phys. Rev. {\bf 135}, A550 (1964).
\bibitem{lo65} A. I. Larkin and Y. N. Ovchinnikov, 
Sov. Phys. JETP {\bf 20}, 762 (1965).
\bibitem{MiMaI} T. Mizushima
{\it et al.}, Phys. Rev. Lett. {\bf 94} 060404 (2005).
\bibitem{SheehyRadzihovskyPRL} D. Sheehy and L. Radzihovsky, Phys. Rev. Lett. {\bf 96}, 060401 (2006).
\bibitem{SheehyRadzihovskyAOP} D. Sheehy and L. Radzihovsky, 
Ann. Phys. (N. Y.) {\bf 322}, 1790 (2007).
\bibitem{commentFRequalmass} For unequal masses FR interaction is
  given by $H_{FR}=\alpha\hat{\bm\phi}^{\dag}\cdot
  \left[\frac{m_1}{m_1+m_2}\hat\psi_{1}(-i\grad)\hat\psi_{2}
    -\frac{m_2}{m_1+m_2}\hat\psi_{2}(-i\grad)\hat\psi_{1}
  \right]+H.c.$ as required to preserve Galilean invariance. L.R.
  thanks V. Gurarie for discussion on this point.
\bibitem{RegalPwave} C. A. Regal
{\it et al.}, Phys. Rev. Lett. {\bf 90}, 053201 (2003).
\bibitem{ChoiRadzihovskyPRA} S. Choi and L. Radzihovsky (to be published).
\bibitem{Greene} B. D. Esry
{\it et al.}, Phys. Rev. Lett. {\bf 78}, 3594 (1997). 
\bibitem{LiuFisher} K.-S. Liu and M. E. Fisher, J. Low
  Temp. Phys. {\bf 10}, 655 (1973).
\bibitem{KNF} J. M. Kosterlitz
{\it et al.}, Phys. Rev. B 13, 412 (1976).
\bibitem{Vicari} P. Calabrese 
{\it et al.}, Phys. Rev. B {\bf 67}, 054505 (2003).
\bibitem{HoSpinor} T.-L. Ho, Phys. Rev. Lett. {\bf 81}, 742 (1998).
\bibitem{Ohmi} T. Ohmi and K. Machida, J. Phys. Soc. Jpn. {\bf 67}, 1822 (1998).
\bibitem{StamperKurn} D. M. Stamper-Kurn
{\it et al.}, Phys. Rev. Lett. {\bf 80}, 2027 (1998).
\bibitem{Cornell} M. R. Matthews
{\it et al.}, Phys. Rev. Lett. {\bf 81}, 243 (1998).
\bibitem{Mukerjee} S. Mukerjee, C. Xu, and J. E. Moore,
  Phys. Rev. Lett. {\bf 97}, 120406 (2006).
\bibitem{commentU_N} $U_N(1)$ and $U_{\Delta N}(1)$ symmetries are associated with total
  atom number $N=N_1+N_2 + 2N_m$ and difference
  $\Delta N = N_1-N_2$ conservations, respectively.
\bibitem{singleQcomment} In the single ${\bf Q}$ condensation
  (Fulde-Ferrell type\cite{ff64}), the state (e.g., atom density) is
  uniform and the phase $\theta$ of $\Psi_-$ is nothing but
  the superfluid phase. For the state with at least two $\pm{\bf Q}$
  condensation (Larkin-Ovchinnikov type\cite{lo65}), the two phases,
  $\theta_{\pm{\bf Q}}$ determine the superfluid phase
  $\theta=(\theta_{\bf Q}+\theta_{-{\bf Q}})/2$ and the density wave
  phonon $u=\frac{1}{2}(\theta_{\bf Q}-\theta_{-{\bf
      Q}})/Q$\cite{RadzihovskyVishwanathPRL}.  
\bibitem{RadzihovskyVishwanathPRL} L. Radzihovsky and A. Vishwanath,
 Phys. Rev. Lett. {\bf 103}, 010404 (2009). 
\end{thebibliography}

\end{document}